\documentclass{article}
\usepackage{arxiv}
\usepackage[utf8]{inputenc} 
\usepackage[T1]{fontenc}    
\usepackage{hyperref}       
\usepackage{url}            
\usepackage{booktabs}       
\usepackage{amsfonts}       
\usepackage{nicefrac}       
\usepackage{microtype}      
\usepackage{lipsum}		    
\usepackage{amsmath,amssymb,amsthm}
\usepackage{bbm}
\usepackage{doi}
\usepackage{graphicx}
\usepackage{natbib}
\usepackage{lineno}
\usepackage{xspace}
\graphicspath{{figures/}}
\hypersetup{pdfborderstyle={/S/U/W 1}}
\newcommand{\orcid}[2]{\href{https://orcid.org/#1}{\includegraphics[scale=0.07]{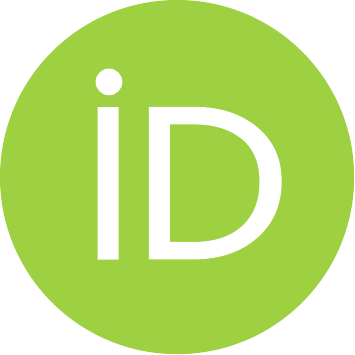}\,#2}}
\newcommand{\ourmethod}{\textit{VMF}\xspace}
\newcommand{\etal}{{\it et al.}\xspace}
\renewcommand{\cite}[1]{\citep{#1}}

\title{Convolutional module for heart localization and segmentation in MRI}

\author{%
\orcid{0000-0002-7818-6103}{D. Lima (1, 2)}\thanks{Corresponding author: danielm@usp.br},
\orcid{0000-0001-7274-2044}{C. Graves (2)},
\orcid{0000-0003-0964-6222}{M. Gutierrez (2)},
\orcid{0000-0001-6167-8104}{B. Brandoli (3)},
\orcid{0000-0001-8318-1780}{J. Rodrigues-Jr (1)},
\vspace{3pt}\\
    (1) ICMC-USP -- Institute of Mathematical and Computer Sciences, University of Sao Paulo\\
	Avenida Trabalhador Sao-carlense 400, 13566-590, Sao Carlos, SP, BR
\vspace{3pt}\\
	(2) Biomedical Informatics Lab, InCor/HC.FMUSP -- Heart Institute, University of Sao Paulo\\
	Avenida Doutor Eneas de Carvalho Aguiar 44, 05403-900, Sao Paulo, SP, BR
\vspace{3pt}\\
	(3) Dalhousie University, 1459 Oxford Street, B3H 4R2, Halifax, NS, CA}

\date{}

\renewcommand{\headeright}{{arXiv} Preprint}
\renewcommand{\undertitle}{{arXiv} Preprint}
\renewcommand{\shorttitle}{Convolutional module for heart localization and segmentation in MRI}

\hypersetup{
pdftitle={Convolutional module for heart localization and segmentation in MRI},
pdfauthor={D. Lima et al.},
}

\begin{document}
\maketitle

\begin{abstract}
Magnetic resonance imaging (MRI) is a widely known medical imaging technique used to assess the heart function. Deep learning (DL) models perform several tasks in cardiac MRI (CMR) images with good efficacy, such as segmentation \cite{Bernard2018}, estimation \cite{xue2018full}, and detection of diseases \cite{khened2017densely}.
Many DL models based on convolutional neural networks (CNN) were improved by detecting regions-of-interest (ROI) either automatically or by hand.
In this paper we describe \emph{Visual-Motion-Focus} (\ourmethod), a module that detects the heart motion in the 4D MRI sequence, and highlights ROIs by focusing a Radial Basis Function (RBF) on the estimated motion field.
We experimented and evaluated \ourmethod on three CMR datasets, observing that the proposed ROIs cover 99.7\% of data labels (Recall score), improved the CNN segmentation (mean Dice score) by 1.7 ($p<.001$) after the ROI extraction, and improved the overall training speed by 2.5 times (+150\%).
\end{abstract}

\section{Introduction}
\label{sec:intro}

Magnetic resonance imaging (MRI) is a medical imaging technique used to capture volumetric image sequences of internal soft tissues, such as cardiac muscles.
In comparison to X-Ray imaging (XR) and Computer Tomography (CT), MRI provides images with improved structural details via finer spatial resolutions.
Cardiac MRI (CMR) focuses on the heart, allowing trained cardiologists to measure heart parameters, for example the mass of the cardiac muscle (myocardium mass), the volumes of blood cavities (atrial and ventricular volumes) and the amount of blood pumped per heartbeat (ejection fraction) \cite{peng2016}.
Those parameters are used to assess how healthy is the heart, by recognizing early conditions and signs before the onset of infarcts and other complications.

Due to the size and complexity of CMR sequences, complex techniques are required to produce detailed analyses; one of these techniques is deep learning (DL). Many of the tasks and goals related to the cardiac functional analysis -- for example segmentation of structures \cite{Bernard2018}, estimation of heart parameters \cite{xue2018full}, and detection of diseases \cite{khened2017densely} -- have benefited from DL methods.
For even better results, research in DL has pointed out that models based on convolutional neural networks (CNN) have had a higher efficacy when provided with regions-of-interest (ROI) either explicitly or implicitly \cite{xue2018full}. The detection of ROIs, usually named {\it ROI proposal}, is a preprocessing step whose goal is to identify the most prominent regions of an image (frame) for discovering clinically-relevant artifacts.

The explicit ROI proposal approaches usually follow a combination of methods, for example:
(a) pipelining a segmentation and a regression network; (b) preprocessing the input with a region proposal algorithm \cite{he2015spatial} or with a CNN \cite{wu2020lv}; or (c) by using manual cropping \cite{xue2017full}.
The implicit ROI detectors are added to the DL network abstracted as additional operators and variables; e.g. multi-scale ``Inception'' \cite{szegedy2015going} and attention \cite{vaswani2017attention} modules, which benefit from the ROIs to down-weight less-informative neurons and inputs inside the network. Inception modules weight convolutions of different sizes, while attention modules assign a weight to each feature channel. This additional neural-network information processing guides which input features or channels shall have more weight so to improve the accuracy of the output layer.

In this paper we develop a module that highlights regions in the image sequence by analyzing the motion field using a Radial Basis Function (RBF).
In our experiments we analyze our method by using the RBF for cropping the input before having it processed by a pretrained segmentation CNN.
Our methodology is an innovation in the task of region proposal for CMR analysis; as we demonstrate, we achieved results that justify the use and further investigation of the employed principles. We named it after its working mechanism as \ourmethod~-- Visual-Motion-Focus.
Section~\ref{sec:prelim} presents the theory and related works regarding ROI proposal and motion detection.
Section~\ref{sec:method} details \ourmethod, presenting its architecture and describing how each step works.
Section~\ref{sec:exps} shows the experiments over \ourmethod, where we demonstrated excellent performance for cropping the ROI, according to metric \emph{Recall}.
Additionally, we also report our results regarding both \emph{Dice} and \emph{Speed-up} metrics.
In conclusion, this work demonstrates the application of technique \ourmethod in objective tasks and how to achieve such a setting.
Section~\ref{sec:conclusion} discusses conclusions and future works.

\section{Theory and related work}
\label{sec:prelim}

\subsection{Cardiac MRI}

MRI is the most precise medical imaging technique for examination of the heart structures, it allows the recording of heart images along a complete heartbeat cycle \cite{earls2002cardiac,seraphim2020quantitative}. In practice the magnetization signal is triggered by a reference pulse, then captured several times for noise reduction and, finally, reconstructed by inverse FFT. The resulting image is usually visualized in slices along a positional axis: long-axis has a frontal or lateral view of the heart, and short-axis aligns to a cross-sectional plane.

\begin{figure}[htb]
    \centering
    \includegraphics[width=0.22\columnwidth]{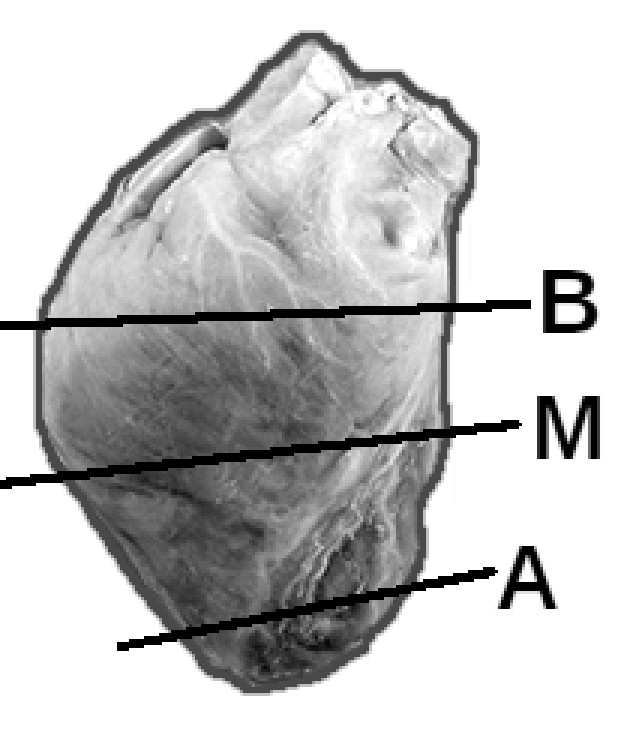} ~
    \includegraphics[width=0.22\columnwidth]{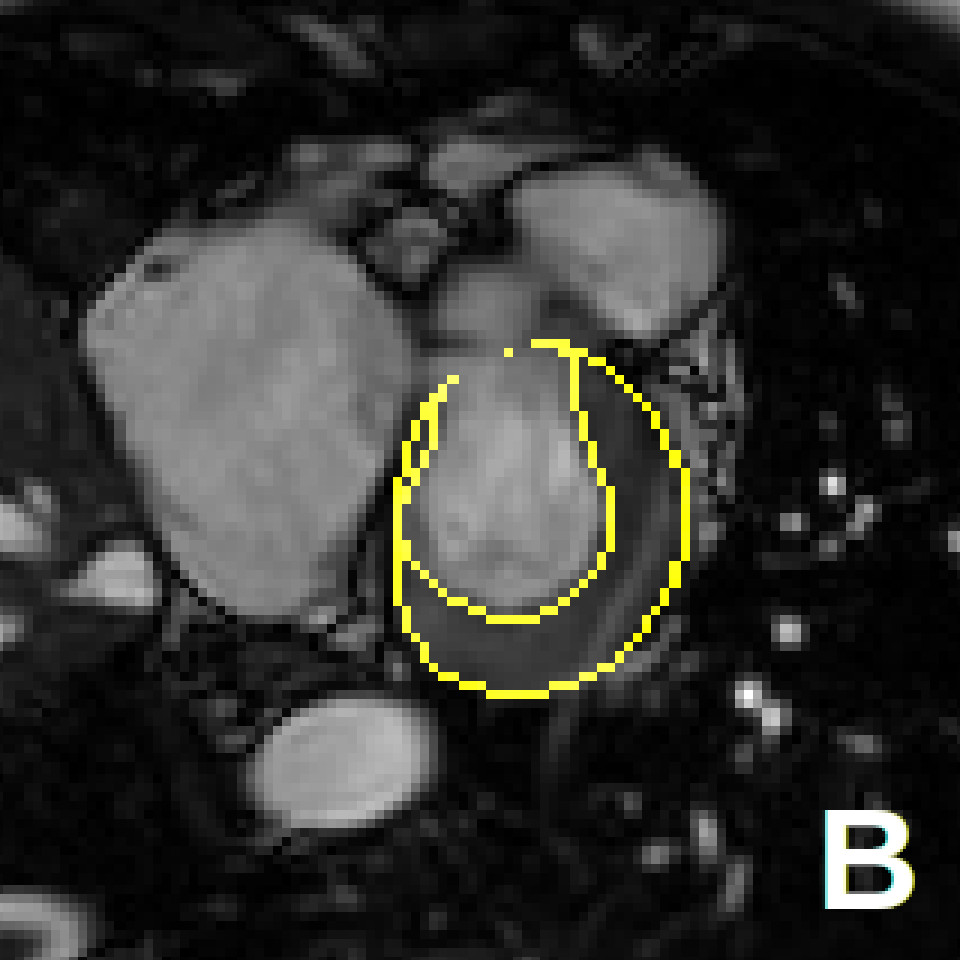} ~
    \includegraphics[width=0.22\columnwidth]{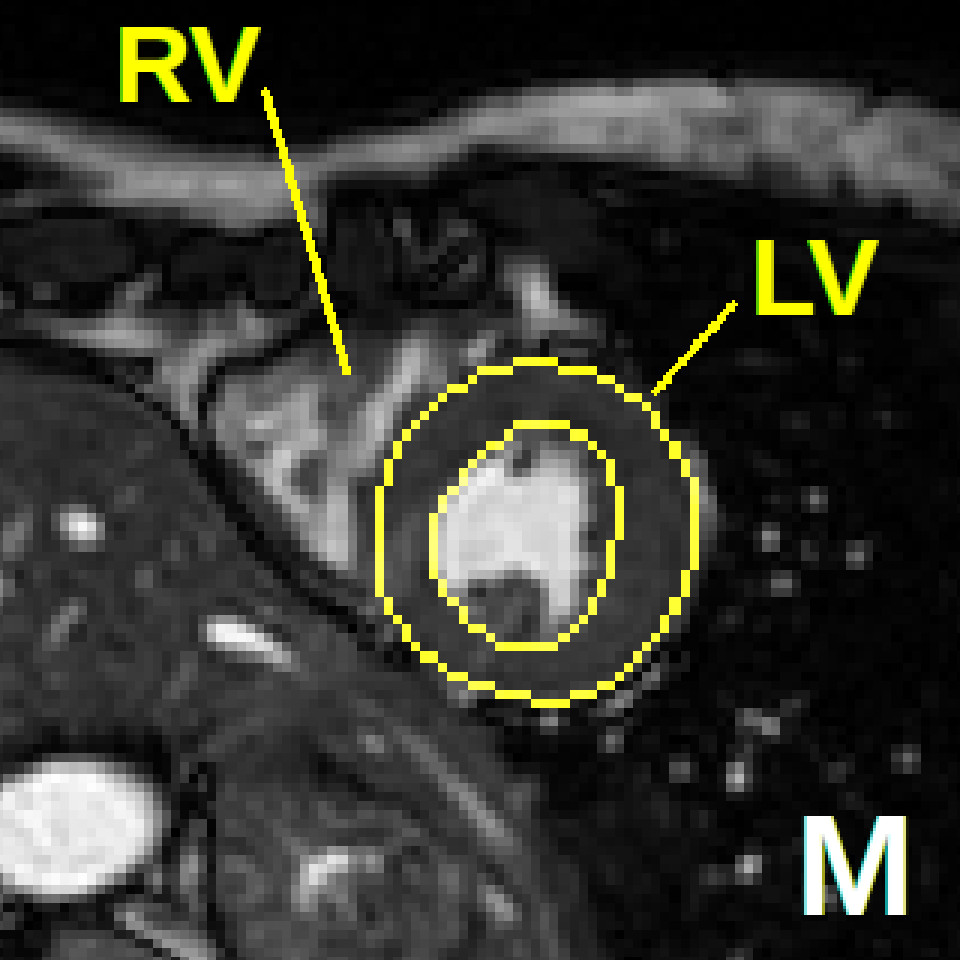} ~
    \includegraphics[width=0.22\columnwidth]{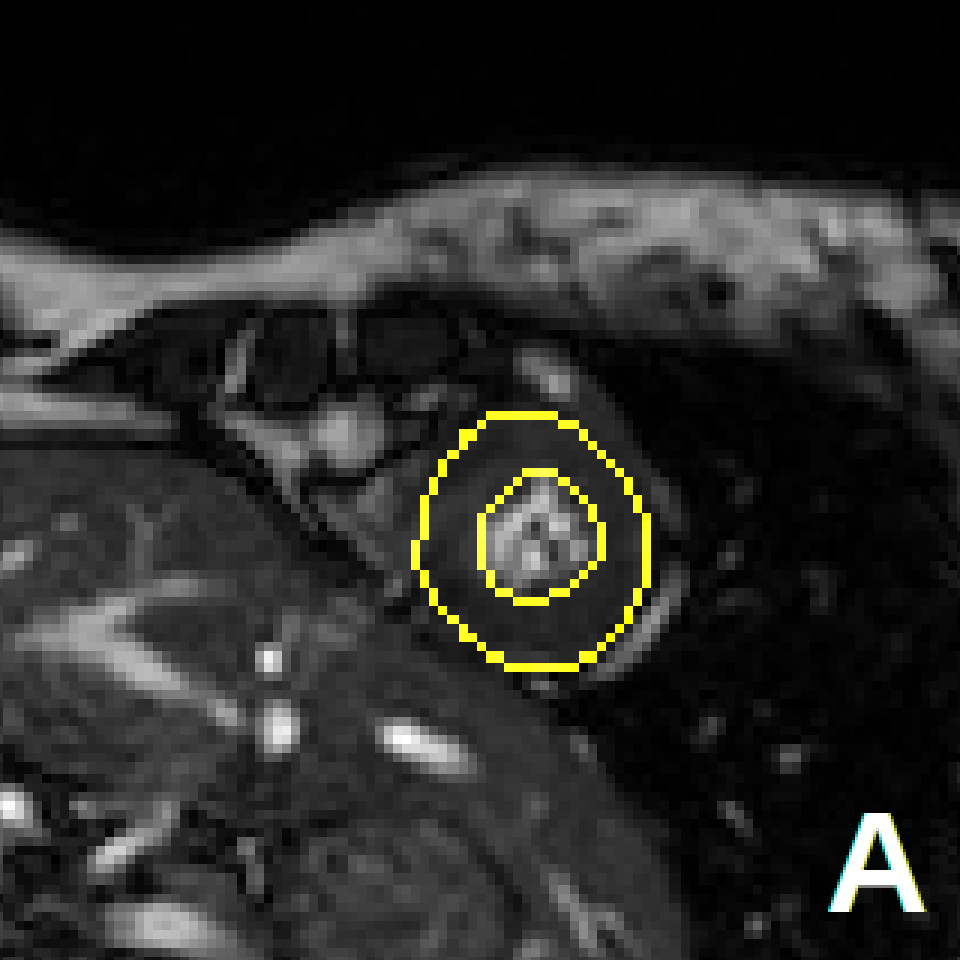}
    \caption{Picture of heart, and examples of short-axis CMR slices in the Base, Middle and Apex regions. The medial slice shows the right and left ventricles (RV and LV). LV boundaries are highlighted in the three slices -- epicardium is the outer boundary and endocardium is the inner boundary. Pictures adapted from the datasets used in \cite{iaizzo2013importance,saber2020multi}.}
    \label{fig:saxcmr}
\end{figure}

The short-axis view is split in three regions: the base or basal region near the top where blood vessels connect to the heart (slice B); the middle or medial in between (slice M); and the apex or apical region at the bottom tip of the heart (slice A) -- refer to Figure \ref{fig:saxcmr}.
The normal human heart has four chambers: right atrium (RA), right ventricle (RV), left atrium (LA) and left ventricle (LV). The atria receive blood and the ventricles pump it out of the heart. Even though all chambers are important, the LV is of special interest because it is the cardiac muscle that does the ``heavy lifting'' of pumping oxygenated blood from the lungs to the whole body. In the short-axis view, the LV appears as a ring shape, whose thickness and internal volume measurements are essential to estimate the myocardium mass and ejection fraction, respectively.

\subsection{Region proposal and Image segmentation}

Region proposal is at the core of image analysis; it is responsible for determining the direction to where, in the image, the analytical algorithms should focus at.
Region proposal methods range from classical algorithms (statistical features, signal processing, binary morphology, etc.) to very complex models such as modern Convolutional Neural Networks (CNNs).
The goal of such algorithms varies from small outputs such as center points or bounding boxes (e.g. quad-tree segmentation, R-CNN and YOLO \cite{spann1985quad,liu1994multiresolution,girshick2014rich,redmon2016you}) to fine contours (e.g. active contour models, region-growing segmentation, U-net and Mask-RCNN \cite{ranganath1995contour,muhlenbruch2006global,he2017mask}).


In many methodologies, the detected ROIs are the starting point of supervised learning with labeled data. However, region-proposal is also organized according to the region under analysis, e.g. image-level, polygon-level, and pixel-level. In this paper we use the nomenclature from \cite{lin2014microsoft}: (a) \emph{Image classification} assigns labels to the whole image (all of its objects); (b) \emph{Object identification} assigns labels to the bounding box region of the detected objects; (c) \emph{Semantic segmentation} assigns labels in polygonal or pixel-level regions, producing finer contours; (d) \emph{Instance segmentation} not only detects the objects' regions but also each individual instance of a detected label. Thus, in order to clear any confusion, we consider that \emph{parameter estimation} is a regression task, and that \emph{image segmentation} is a polygon or pixel-level classification task as mentioned above.

\subsection{Image features and Motion analysis}

Some diseases may decrease the extension and intensity of heart movements; in such cases, the motion features will have little to no information. A way to address this issue is by combining the static visual features of the image and the dynamic features. The static visual features are computed by classical feature extractors, then are adjusted to detect the heart that, in short-axis CMRs, appears as a gray bubble with dark borders. The dynamic features are computed by motion estimation.

Many categories of image feature extractors are present in the literature \cite{gonzalez2002digital}, such as statistical moments (mean, standard deviation, skewness, kurtosis), robust statistics (median, quartiles, interquartile range, percentiles, rank, median absolute error), signal processing filters (Gaussian, Sobel, Laplacian, Gabor) and transforms (Hough, FFT, DCT, DWT), among other measures of information distribution (Shannon entropy, Minkowski-Bouligand box-counting dimension, Hausdorf fractal dimension).

Motion estimation refers to a set of techniques to explore the temporal redundancy in sequences of images, with applications in video compression, analysis and processing \cite{Oliveira2006}. One classic formulation of motion analysis is optical flow \cite{horn1981determining}, which estimates the displacement field of an object by smooth derivatives of brightness, assuming that the brightness of moving objects does not change between frames. Other approaches include statistical and block-based algorithms; change detection by individual or accumulated frame difference; and parametric methods \cite{Oliveira2006}. Motion estimation is mainly used for general 2D video coding, camera motion compensation \cite{thanou2015graph,zhang2018fast,bao2019memc}, and for medical imaging \cite{stemkens2016image,mohsin2017accelerated,yan2018left,yu2020foal}.

\subsection{Radial Basis Functions}

A Radial Basis Function (RBF) is a real-valued function of the radius (or distance) from an origin to a given point \cite{majdisova2017radial}. They have several applications in function approximation and neural networks \cite{broomhead1988radial,buhmann2003radial}.
RBFs can be understood as a flashlight beam on a wall, which is intense at the center and fades sideways. RBF kernels are built by computing the distance of each pixel position from the center, then transforming this distance by a decaying function (e.g. exp, log, Gaussian).
Many center functions can be used as the RBF center, such as statistical measures of central tendency (e.g. simple and weighted mean, median, center-of-mass). Examples of distance functions include the standard deviation (also known as Malahanobis distance, or z-score difference), Minkowski (e.g. $L_1$, $L_2$, $L_\infty$), Jaccard, and Cosine distances.

\subsection{Computer methods for CMR analysis}

Computer methods for functional analysis of CMR as reviewed by Peng \etal \cite{peng2016} were organized in three ways: image-driven, model-driven, and by direct estimation. Further subdivisions of the LV segmentation spans five groups: (1) image processing methods such as thresholding, morphology operators, and region-growing; (2) pixel/voxel-based classification by gaussian mixture models, neural networks, k-means, k-nearest neighbors, or support vector machines (SVM); (3) active contours (snakes), deformable models, level sets, and motion tracking; (4) principal or independent component analyses (PCA and ICA) with strong priors from statistical models of the heart anatomy; and (5) direct estimation by, e.g., latent discriminant analysis (LDA) combined with SVM. This work refers to category 2 as it employs a pixel-based classification mediated by a neural network.

\subsection{Deep learning region proposal and LV segmentation}
\label{sec:dl_lv}
Recent approaches for LV segmentation use CNN models (such as U-net) experimented over diverse datasets and methodological combinations. U-net \cite{ronneberger2015u} is a general segmentation model which combines a tower of downscaled-then-upscaled deep representations by concatenation with the input features, instead of addition like FPN \cite{lin2017feature} and LinkNet \cite{chaurasia2017linknet}, with success in LV segmentation. The approach introduced by Smistad \etal \cite{smistad2017} departs from the pretraining of an U-net CNN using a Kalman-filter segmentation; it achieves 86\% Dice (epicardium) in ultrasound images -- Dice (DSC) will be detailed in Section \ref{sec:metrics}. For CMR, U-net displayed even better results (89\% Dice) when trained from min-cut priors \cite{guo2018cardiac}. An U-net architecture with residual blocks and optical flow information \cite{yan2018left} achieved 89\%, 95\% and 85\% Dice in the base, middle and apex regions of the heart respectively.
In the work of Wu \etal \cite{wu2020lv}, the authors combine a custom CNN for region proposal with an U-net segmentation to achieve 95\% Dice. Overall, the combination of region proposal to U-net had good results in particular datasets, but still has room for improvement when the evaluation generalizes across multiple datasets. 
Different to former works, our methodology, \ourmethod, uses RBFs to propose ROIs that will aid a CNN processing in the task of LV segmentation. As we demonstrate, we achieve superior results over an ample set of experiments.

\section{Material and methods}
\label{sec:method}
\ourmethod starts with a 4D image input $\mathbf{x} = I(t,x,y,z)$, that is, a sequence of volumes (frames) each one with a time $t$ coordinate. Initially, we normalize $\mathbf{x}$ to produce sequence $\mathbf{x}^*$ with frames in a format more adequate for Neural Network processing -- see Section \ref{sec:imnorm}. From $\mathbf{x}^*$, we extract visual features to produce $\mathbf{x}_s$, and motion estimation to produce $\mathbf{x}_t$, detailed in Section \ref{sec:vismotion}. Next, we apply two sets of weights: $w_s$ the weights related to visual features; and $w_t$, the weights related to the motion, or time; combining both features in tensor $\mathbf{v}$. Then, as presented in Section \ref{sec:centerscale}, we compute the center voxel $\mu_v$ defined by the energy-weighted sum of all the voxels' coordinates; then we produce a segmentation map $\mathbf{y_S}$ by applying a threshold to $\mathbf{v}$ and extracting the bounds of non-zero voxels; afterwards we compute the scale $\sigma_v$ given by the standard deviation of all the voxels' distances from center $\mu_v$. At this point, refer to Section \ref{sec:segloc}, we can apply a Radial Basis Function at center $\mu_v$ with radius $\sigma_v$, computing $\mathbf{y_L}$, then we scale the region defined by $\mathbf{y_L}$ to the CNN input shape. In Section \ref{sec:NNROI}, we explain the Neural Network processing, its parameters and training.

\subsection{Intensity normalization}
\label{sec:imnorm}

We normalized the image intensities between 0 and 1, by subtracting the minimum value and dividing by the range of values, where $\epsilon$ is a small constant to avoid division-by-zero:
\begin{linenomath}\begin{equation}
 \mathbf{x}^* = \frac{\mathbf{x} - \mathrm{min}(\mathbf{x}) + \epsilon}{\mathrm{max}(\mathbf{x}) - \mathrm{min}(\mathbf{x}) + \epsilon}
\end{equation}\end{linenomath}

\subsection{Visual features and motion estimation}
\label{sec:vismotion}

\begin{figure*}
    \centering
    \includegraphics[width=\textwidth]{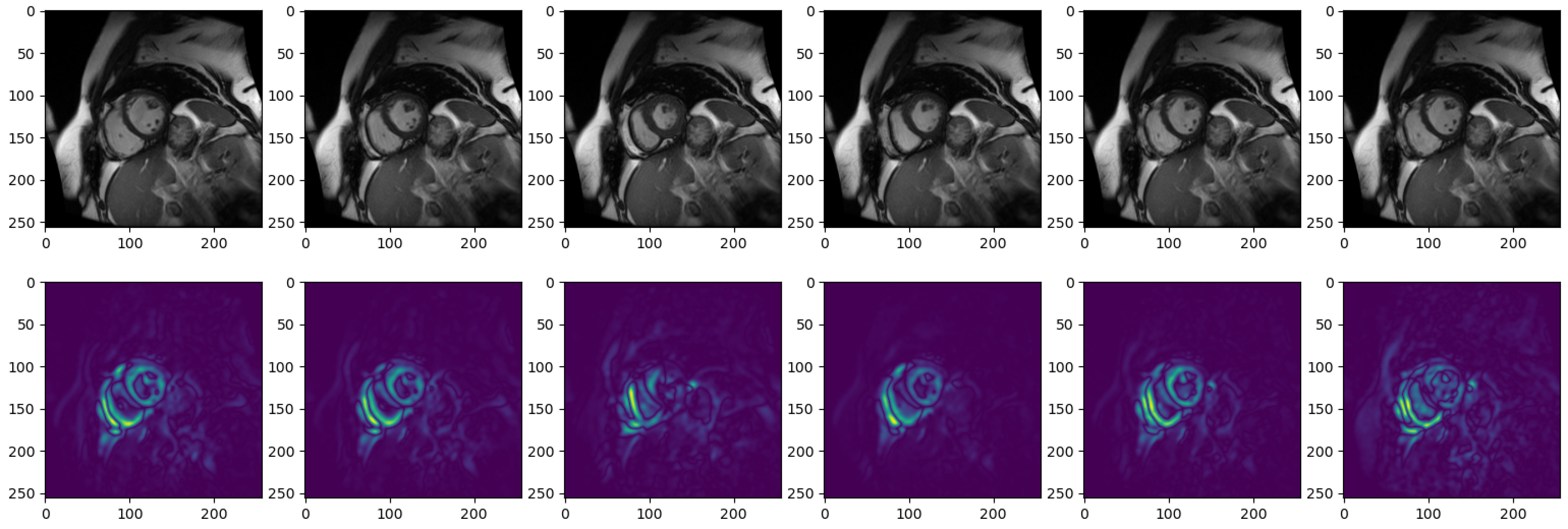}
    \caption{Sequence of images in the medial slice, showing six frames (upper row) and the respective absolute derivatives in the time dimension (lower row). The example in the figure only exhibits the medial slice because it is easier to visualize and interpret, but notice that \ourmethod is defined for volumes.}
    \label{fig:step1}
\end{figure*}

For visual features extraction, we consider the statistical mean and standard deviation obtained with 3 $\times$ 3 $\times$ 3 kernels. 
The mean image $I_\mu$ is computed using the convolution operation with mean kernel $M$, defined as:
\begin{linenomath}\begin{equation}\label{eq:meankernel}
M = \frac{1}{1 \cdot 3 \cdot 3 \cdot 3} \cdot J_{1,3,3,3}\\
\end{equation}\end{linenomath}

\noindent{where $J_{1,3,3,3}$ is a 1 $\times$ 3 $\times$ 3 $\times$ 3 matrix-of-ones. That is, $M$ is just the arithmetic average of a 3 $\times$ 3 $\times$ 3 matrix arranged for convolution. With kernel M, we compute $I_\mu$ as:}
\begin{linenomath}\begin{equation}\label{eq:img_mean}
I_\mu = I \ast M
\end{equation}\end{linenomath}

In turn, the standard deviation image $I_\sigma$ is computed by taking the differences between the image $I$ and mean image $I_\mu$, then squaring the differences element-wise (Hadamard power), after that by convolving with the mean kernel $M$, and then taking the Hadamard square-root, as follows:
\begin{linenomath}\begin{equation}\label{eq:xs}
I_\sigma = [ (I - I_\mu)^{(2)} \ast M ]^{(1/2)}
\end{equation}\end{linenomath}
%

Accordingly, the mean image $I_\mu$ and the standard deviation image $I_\sigma$ define $\mathbf{x}_s = \{ I_\mu, I_\sigma \}$.
Notice that we express the Hadamard powers using the definitions and notations defined in the work of Fallat and Johnson \cite{fallat2007hadamard}.

We used the motion estimate function $E(I)$ -- refer to Equation \ref{eq:xt}, given by the root-mean-squared differences of intensity along the time coordinate, where $T$ is the time interval (or number of frames) in image $I$, and $S_t$ is the Sobel kernel w.r.t. time, instead of the default $S_x$ and $S_y$ spatial Sobel kernels.
This function is related to the magnitude of the optical flow \cite{horn1981determining} vector field in each voxel, as follows:
\begin{linenomath}\begin{equation}\label{eq:xt}
\mathbf{x}_t = E(I) = \sqrt{\frac{1}{T} \int_{t=0}^{T} \left( \frac{\partial I}{\partial t} \right)^2 dt} \approx \sqrt{\frac{1}{T} \sum_{t=0}^{T-1} [I(t) \ast S_t]^{(2)}}
\end{equation}\end{linenomath}

\begin{figure}[htb]
    \centering
    a) \includegraphics[height=.33\columnwidth]{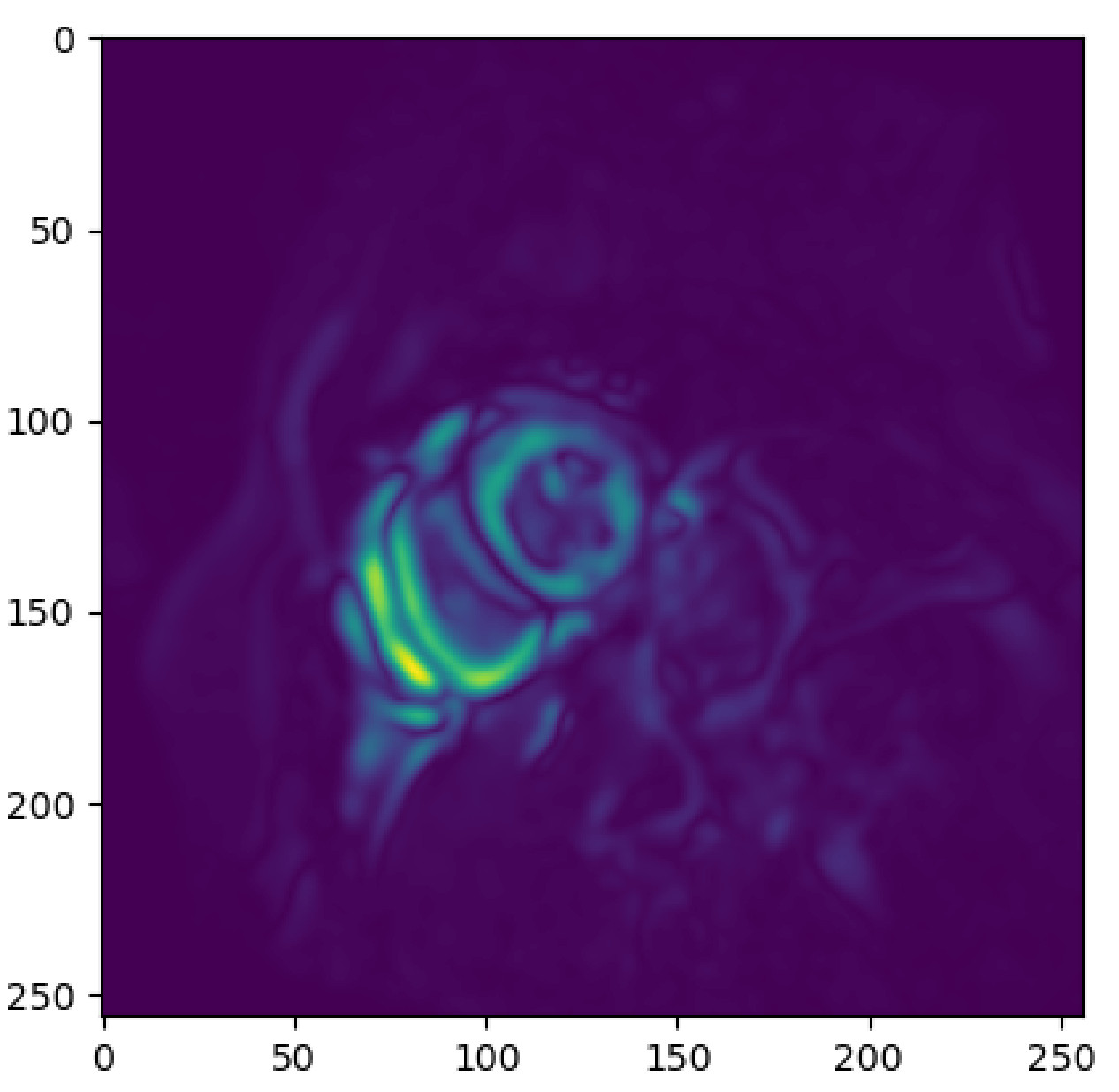}
    \raisebox{.03\columnwidth}{\includegraphics[height=.27\columnwidth]{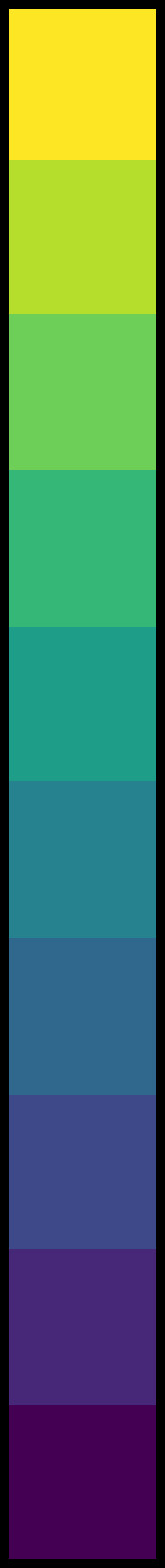}}{\kern-1em\scriptsize $E(I)$} ~ 
    b) \includegraphics[height=.31\columnwidth]{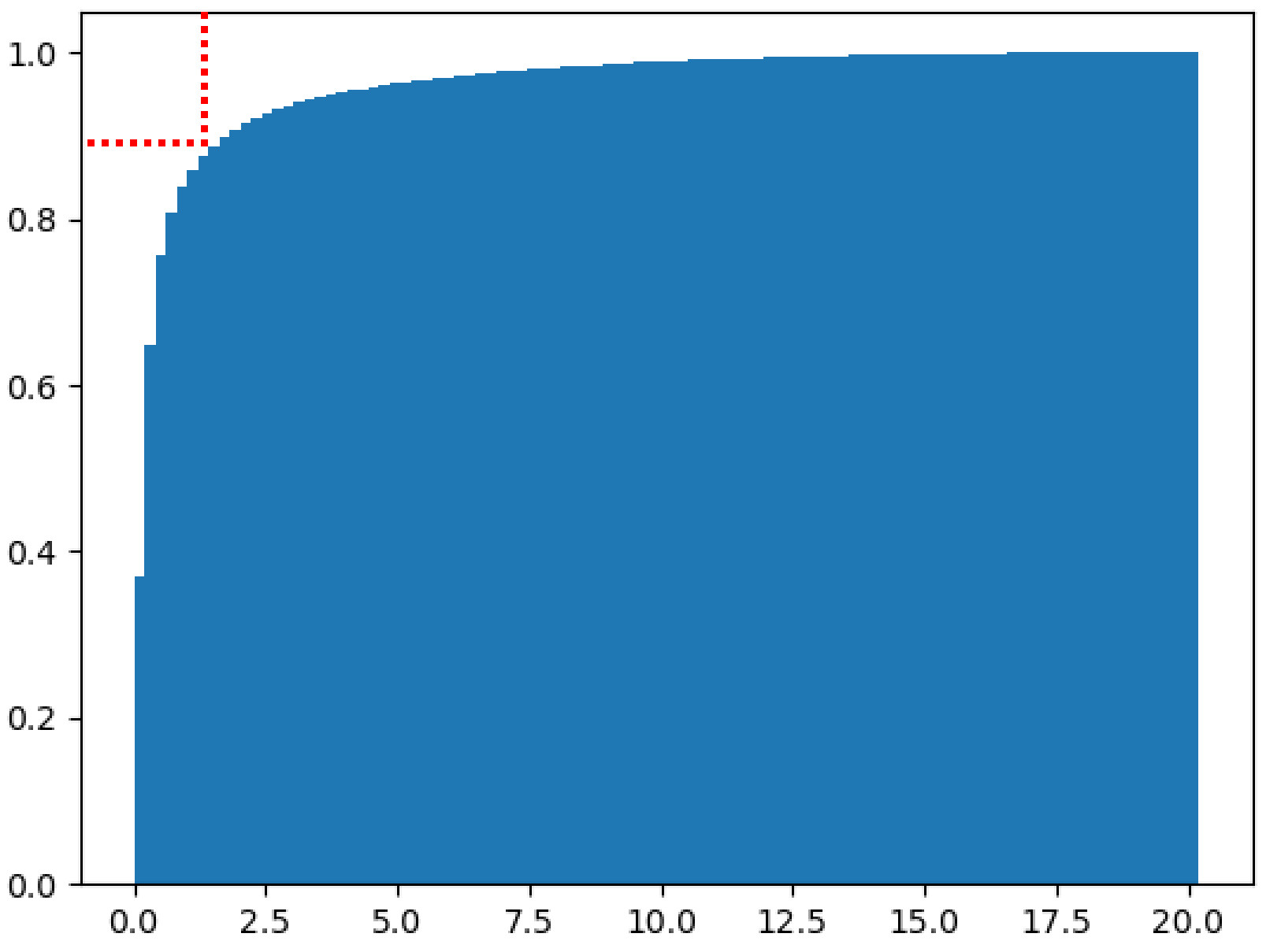}
    \caption{a) Motion estimate $E(I)$ along the whole cardiac cycle. b) Cumulative histogram of the computed values. The histogram shoulder lies near the percentile $p=0.9$, which is used for thresholding.}
    \label{fig:step2}
\end{figure}

Figure \ref{fig:step1} illustrates a sequence of six CMR frames spaced by $1/6$ of the cardiac cycle (upper row in the figure), and the respective absolute derivatives in each point along the time dimension (lower row) as computed with Equation \ref{eq:xt}.
Figure \ref{fig:step2}a shows the total motion estimate in the sequence, while Figure \ref{fig:step2}b presents the cumulative energy histogram. By combining both static and motion features $\mathbf{x}_s$ and $\mathbf{x}_t$, we obtain:
\begin{linenomath}\begin{equation}\label{eq:v}
\mathbf{v} = w_s \mathbf{x}_s + w_t \mathbf{x}_t
\end{equation}\end{linenomath}

The weights $w_s$ and $w_t$ are initialized to $0.1$ and $0.9$ respectively for $\mathbf{x}_s$ and $\mathbf{x}_t$, i.e., although we emphasize the motion features, we also include the static visual features, which will address the problematic cases when the heart has limited motility, which might be the case for heart complications.

\subsection{Center and scale computation}
\label{sec:centerscale}
The center of energy $\mu_v$, formalized by Equation \ref{eq:center}, is the voxel defined by the energy-weighted sum of all the voxel coordinates $\mathbf{r}$ in dimensions $x, y, z$; that is, every voxel $i$ has $\mathbf{r}_i = \langle x_i, y_i, z_i \rangle \ |\ 0 \leq x_i < \mathrm{width}, 0 \leq y_i < \mathrm{height}, 0 \leq z_i < \mathrm{slices}, 1 \leq i \leq N $, for $N$ is the number of voxels in the image, as follows:
\begin{linenomath}\begin{equation}\label{eq:center}
\mu_v = \langle \bar x, \bar y, \bar z \rangle = \frac{\sum_{i=1}^{N} \mathbf{v}_i \cdot \mathbf{r}_i}{\sum \mathbf{v}}
\end{equation}\end{linenomath}

The segmentation $\mathbf{y_S}$ is given by the thresholded image from the previous step, see Figure \ref{fig:step3}. The scale estimate $\sigma_v$ is the cube root of the volume (in voxels) of the thresholded image considering the values above the 90\% percentile:
\begin{linenomath}\begin{align}
\label{eq:ys} (\mathbf{y_S})_i &= \mathbbm{1}[\mathbf{v}_i > Q(0.9, \mathbf{v})] \\
\label{eq:scale} \sigma_v &= \frac{3}{\| \mathbf{r}_{\max} \|} \sqrt[3]{ \sum_{i=1}^{N} \nolimits (\mathbf{y_S})_i }
\end{align}\end{linenomath}

\noindent where the indicator function $\mathbbm{1}[c]$ returns $1$ if the condition $c$ is true, or $0$ otherwise; and $Q(p,v)$ is the quantile function, returning the maximum of the lowest $p$ (\%) values in $v$.

\subsection{Segmentation and localization}
\label{sec:segloc}
The segmentation $\mathbf{y_S}$ is derived from thresholded $\mathbf{v}$ (Equation \ref{eq:ys}).
The localization $\mathbf{y_L}$ is found by fitting an RBF to $\mathbf{y_S}$. The radius $d_i$ is the distance from the center $\mu_v$ to each voxel $i$. The Euclidean distance ($L_2$-norm $\| \cdot \|$) was used:
\begin{linenomath}\begin{equation}
d_i = \left\| \frac{\mathbf{r}_i - \mu_v}{\mathbf{r}_{\max}} \right\| = \sqrt{\left(\frac{x_i - \bar x}{x_{\max}}\right)^2 + \left(\frac{y_i - \bar y}{y_{\max}}\right)^2 + \left(\frac{z_i - \bar z}{z_{\max}}\right)^2}
\end{equation}\end{linenomath}

The chosen RBF is a Gaussian $\phi$ of the voxel distances:
\begin{linenomath}\begin{equation}\label{eq:gauss}
(\mathbf{y_L})_i = \phi(i) = \exp \left[-(d_i / \sigma_v)^2 \right]
\end{equation}\end{linenomath}

\begin{figure}[htb]
    \centering
    a) \includegraphics[height=.41\columnwidth]{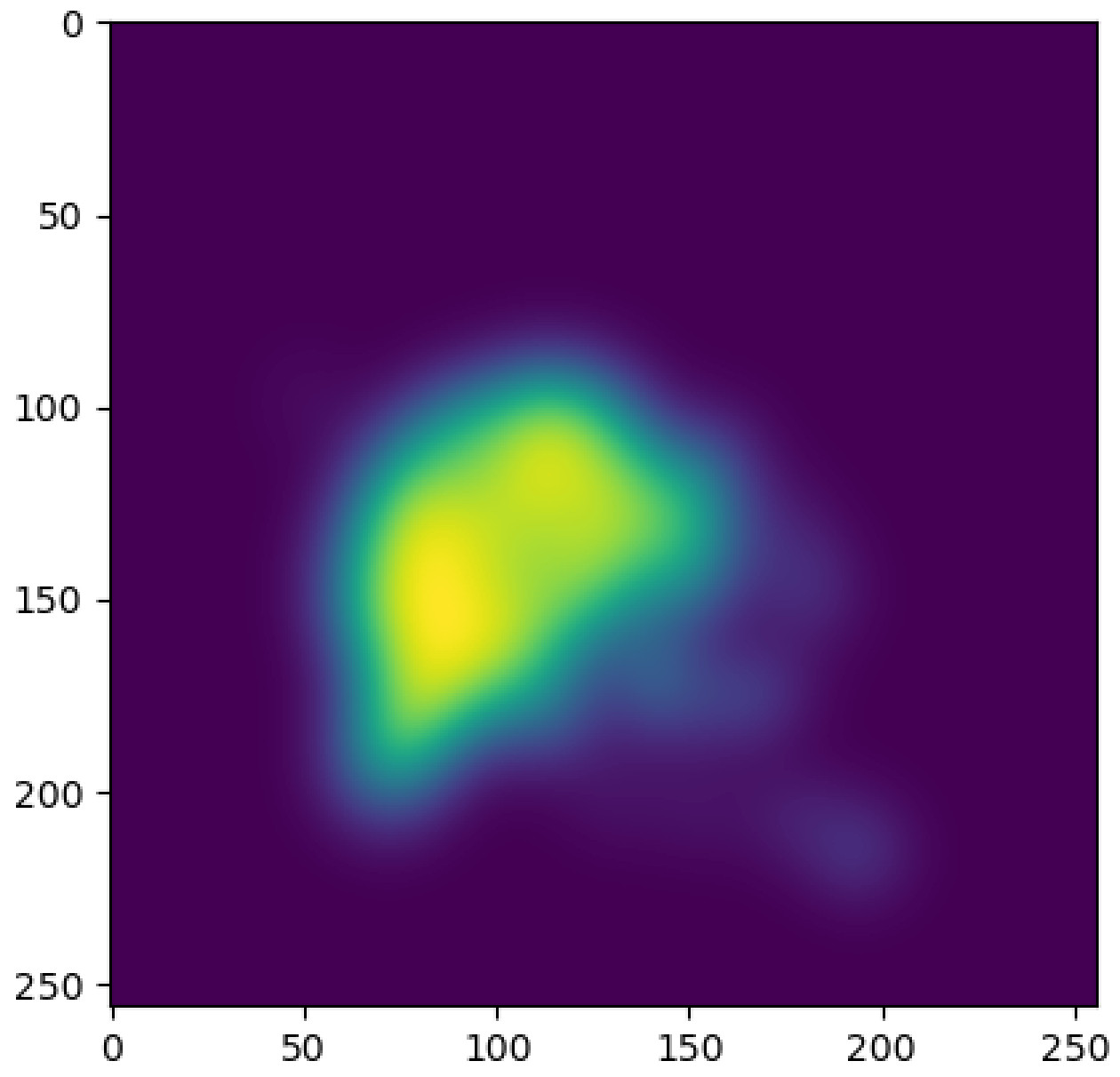} ~ 
    b) \includegraphics[height=.41\columnwidth]{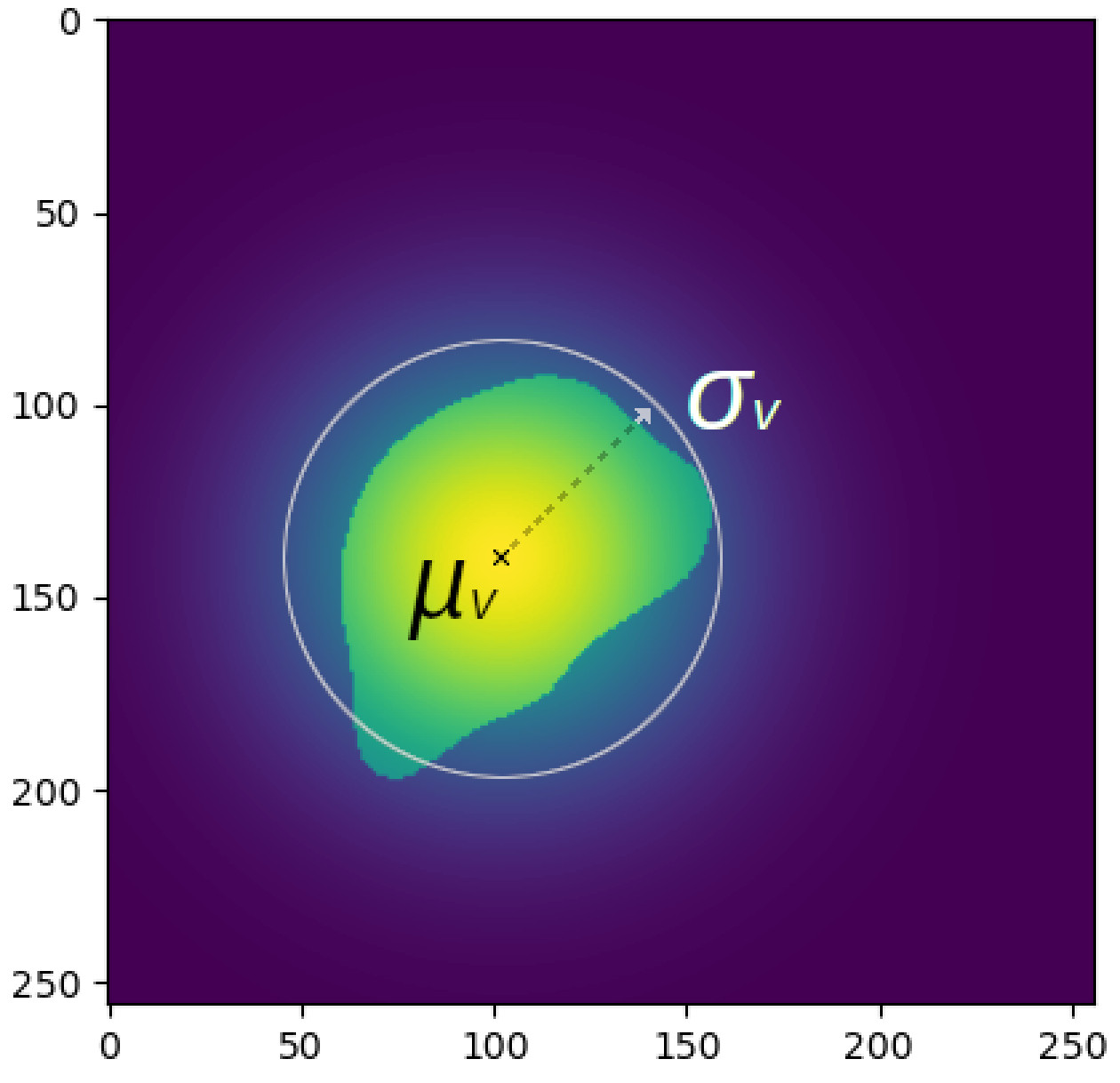}
    \caption{a) $E(I)$ above 90\% of Figure \ref{fig:step2}a, convolved by Gaussian kernel with $\sigma=5$. b) The RBF is fitted to this region with center $\mu_v$ and radius $\propto \sigma_v$.}
    \label{fig:step3}
\end{figure}

Both outputs $\mathbf{y_S}$ and $\mathbf{y_L}$ are illustrated in Fig. \ref{fig:step3}.
According to the framework depicted so far, we can design a focus for many different objects in the images by changing the functions for image feature extraction, motion estimation, center, scale and RBF. It is also possible to detect multiple objects or objects of complex shapes by a mixture of RBF models.

\subsection{Crop, scaling and fine CNN segmentation}
This step performs the final preparation of the image so that it will fit the CNN input shape; this is necessary because the MRI images may have diverse dimensions. After all the RBF estimation steps executed in the original resolution, the proposed region is cropped then adjusted, followed by an intensity normalization. For CNNs with fixed input dimensions, we rescale images using bicubic spline interpolation. For CNNs with variable input shape, we only adjust the image proportions as requested by the model, e.g. U-net has five max-pooling layers with down-scaling factor 2, which means the input dimensions should be multiples of 2$^5$ in order to have at least 1 pixel in the encoder output.

\subsection{Neural Network ROI proposal}
\label{sec:NNROI}
\ourmethod was developed in combination with a 2D U-net CNN \cite{ronneberger2015u} for simplicity, but it is suited to any other convolutional network. The U-net is a general encoder-decoder segmentation architecture with success in LV segmentation, as explained in Section \ref{sec:dl_lv}.
The CNN was trained with outputs obtained with \ourmethod during 30 epochs using an adaptive momentum optimizer, initial learning rate $\eta$ = 0.001, Nesterov $\beta_1$ = 0.9, $L_\infty$-decay $\beta_2$ = 0.999 and the loss below, which is binary cross-entropy plus DSC score:
\begin{linenomath}\begin{equation}
    \mathcal{L}(y,p) = -[y \cdot \log(p) + (1 - y) \cdot \log(1 - p)] + DSC(y,p)
\end{equation}\end{linenomath}

\noindent where DSC is defined by Equation \ref{eq:dsc} in Section \ref{sec:metrics}.

The \ourmethod pipeline executes in the whole cardiac volume and, after cropping, the cropped ROI is passed to the segmentation CNN, which in this case is the 2D U-net.

\subsection{Datasets}

To evaluate the \ourmethod framework we used three CMR datasets specified across the short-axis orientation. The metadata in the datasets include: a) binary masks for the LV and RV; b) physiological parameters such as myocardium mass, ventricle area, volume, ejection fraction, thickness, and dimensions of structures; c) image acquisition parameters such as spatial resolution (mm), temporal resolution (frames per cardiac cycle) and slice gaps (mm). Not all of the datasets encompass the same information; following we provide more details, with a summary presented in Table \ref{tab:data}.

\begin{itemize}
    \item \textbf{LVSC} - Cardiac Atlas Project (CAP) 2011 LV Segmentation Challenge \cite{SUINESIAPUTRA2014}. We used the 100 patients in the training set, which have LV masks on all the frames;
    \item \textbf{ACDC} - MICCAI 2017 Automated Cardiac Diagnosis Challenge \cite{Bernard2018}, training dataset. It was created from clinical data, including sequences of 100 patients from the University Hospital of Dijon (France) over 6 years, it includes RV and LV masks in two frames; 
    \item \textbf{M\&Ms} - MICCAI 2020 Multi-Centre, Multi-Vendor \& Multi-Disease Cardiac Image Segmentation Challenge \cite{saber2020multi}. This database was collected from six hospitals in Spain, Canada, and Germany using several MRI scanners (Siemens, GE, Philips and Canon), it includes RV and LV masks in two frames. We used 320 patients for which the labels are publicly available.
\end{itemize}

\begin{table}[htb]
    \centering{\footnotesize\sffamily
    \begin{tabular}{lccc}
        \hline
        \textit{\textbf{Dataset}} &\textbf{LVSC} &\textbf{ACDC} &\textbf{M\&Ms}\\
        \hline
        \textbf{n}       & 100       & 100       & 320       \\
        \textbf{width}   & 138-512   & 154-428   & 196-548   \\
        \textbf{height}  & 138-512   & 154-512   & 192-512   \\
        \textbf{slices}  & 8-24      & 6-18      & 6-20      \\
        \textbf{frames}  & 18-35     & 12-35     & 18-36     \\
        \textbf{Sxy}     & 0.68-2.14 & 0.70-1.92 & 0.68-1.82 \\
        \textbf{Sz}      & 6-10      & 5-10      & 5-10      \\
        \hline
    \end{tabular}}
    \caption{Overview of the three CMR datasets: n = number of patients; (width,height,slices) = spatial image dimensions in (x,y,z) pixels; frames = number of images in the time coordinate; Sxy = spatial resolution (pixel spacing) in the axial plane (mm/pixel); Sz = slice resolution (mm/pixel).}
    \label{tab:data}
\end{table}

\section{Experiments}
\label{sec:exps}

In order to evaluate \ourmethod, we analyzed the selected datasets by applying \ourmethod, then compared the results of two CNNs: one base CNN on the raw images without \ourmethod, and another CNN combined with the images processed by \ourmethod.
After the execution of \ourmethod and cropping of ROIs, only the slices and frames with label masks of at least 25 pixels are passed to the U-net (i.e. stray markings are discarded). The labels are obtained by subtracting epicardium and endocardium masks, i.e. they contain the myocardium region. The datasets were used individually for training and validation of models, then combined in an all-versus-all scheme.

\subsection{Metrics}
\label{sec:metrics}

We compared three indices: (1) Recall, the proportion of the labels that was preserved in the \ourmethod region proposals; it is also known as Sensitivity or True Positive Rate -- TPR as defined by Equation \ref{eq:tpr}, which aims to verify if the bounding boxes cover the labels entirely; (2) the Sørensen-Dice Coefficient -- DSC as defined by Equation \ref{eq:dsc}, which is equivalent to the F1-score (average of Precision and Recall) that refers to the segmentation output when not-using vs using method \ourmethod; (3) speedup, the ratio of the times taken by the segmentation CNN when not-using vs using method \ourmethod~-- it is defined as ($t_{\mathrm{base}} / t_{\mathrm{ours}}$).
\begin{linenomath}
\begin{equation}\label{eq:tpr}
TPR(y,\hat{y}) = \frac{|y \cap \hat{y}|}{|y|} = \frac{TP}{TP+FN}
\end{equation}
\begin{equation}\label{eq:dsc}
DSC(y,\hat{y}) = \frac{2|y \cap \hat{y}|}{|y|+|\hat{y}|} = \frac{2TP}{2TP + FP + FN}
\end{equation}
\end{linenomath}

Recall ranges from 0.0, when none of the marked voxels are detected; to 1.0, when 100\% of the marked voxels are detected. DSC ranges from 0.0, for no intersection; to 1.0, for a perfect match between $y$ and $\hat{y}$. Speedup is a positive ratio with value 1.0 when both methods take the same time; $>$1.0 when \ourmethod is faster; and $<$1.0 when \ourmethod is slower.

\subsection{Test Setup}

The experiments ran in a rack-mounted machine with Intel i7-7700k CPU and NVIDIA Titan-Xp GPU. Our scripts were written in Python, scipy, keras/tensorflow, and matplotlib.
The steps were: load an image; compute features; fit RBF; crop and resize both the input image and the label mask; Base CNN prediction on the original image and \ourmethod-CNN prediction on the cropped image; then, evaluate the predicted masks against the labels using the aforementioned metrics.

\section{Results}
\label{sec:results}

Table \ref{tab:results} presents the results of \ourmethod region proposals (Recall) and \ourmethod-CNN combination (Dice score).
After applying method VMF, the U-net CNN trained on dataset ACDC had the worst Dice scores, with a mean performance decrease of 8.3\% when considering all the datasets.
On the other hand, the model trained on dataset M\&Ms was significantly improved when using the \ourmethod ROI proposal, with a mean performance increase of +7.2 (percent points) considering all the datasets.
With dataset LVSC, we observed a lower performance when validating over dataset LVSC (-3.5), but an improvement when validating with datasets ACDC and M\&Ms (+4.0 and +2.5, respectively).

\begin{table}[ht]
    \centering{\footnotesize\sffamily
    \begin{tabular}{ccccr}
    \hline
    \textbf{Train Set} &\textbf{Test} &\textbf{Base} &\textbf{\ourmethod} &\textbf{Dice} \\
    \textbf{(Recall)} &\textbf{Set} &\textbf{Dice} &\textbf{Dice} &\textbf{Diff.}\\
    \hline
              & LVSC  & 50.1\% & 44.8\% & -5.7 \\ 
    ACDC      & ACDC  & 74.6\% & 69.9\% & -4.7 \\ 
    Recall of 98.84\% & M\&Ms & 67.6\% & 47.4\% & -20.2\\ 
              & all   & 56.6\% & 48.3\% & -8.3 \\ 
    \hline
              & LVSC  & 62.4\% & 70.9\% & +8.5 \\ 
    M\&Ms     & ACDC  & 60.8\% & 67.0\% & +6.2 \\ 
    Recall of 99.75\% & M\&Ms & 81.3\% & 82.2\% & +0.9 \\ 
              & all   & 65.4\% & 72.6\% & +7.2 \\ 
    \hline
              & LVSC  & 74.3\% & 70.8\% & -3.5 \\ 
    LVSC      & ACDC  & 75.7\% & 79.7\% & +4.0 \\ 
    Recall of 99.75\% & M\&Ms & 67.0\% & 69.5\% & +2.5 \\ 
              & all   & 74.5\% & 72.7\% & -1.8 \\ 
    \hline
              & LVSC  & 76.5\% & 77.5\% & +1.0 \\ 
    all       & ACDC  & 85.9\% & 86.8\% & +0.9 \\ 
    Recall of 99.69\% & M\&Ms & 77.2\% & 83.0\% & +5.8 \\ 
              & all   & 77.7\% & 79.4\% & +1.7 \\ 
    \hline
    \end{tabular}}
    \caption{Performance results. In the first (left-most) column, the name of the dataset and the Recall metric (Section \ref{sec:metrics}. In the remaining columns, the Dice scores (metric DSC in Section \ref{sec:metrics}) for CNN segmentation without \ourmethod (Base) and combined with \ourmethod.}
\label{tab:results}
\end{table}

We applied the McNemar's chi-squared test with continuity correction from package statsmodels 0.12.2, whose results asserted that the Base predicitons and predictions after \ourmethod were significantly different, with $p<.001$  (see Table \ref{tab:results3}).

\begin{table}[ht!]
    \centering{\footnotesize\sffamily
    \begin{tabular}{crr}
    \hline
    &\textbf{\ourmethod} (T) &\textbf{\ourmethod} (F) \\
    \hline
    \textbf{Base} (T) & 6,041,403 & 2,543,222 \\
    \textbf{Base} (F) & 818,157 & 709,767,778 \\
    \hline & $\chi^2=885305.1$ & $(p<10^{-302})$
    \end{tabular}}
    \caption{Contingency matrix considering all voxels in the configuration train:all versus test:all, refer to Table \ref{tab:results}; and the corrected McNemar's chi-squared test comparing CNN without \ourmethod (Base) versus with \ourmethod asserted significantly different predictions, with $p<.001$.}
    \label{tab:results3}
\end{table}

From a practical perspective we also compared the training speed, which is paramount to larger experiments such as hyper-parameter search. Table \ref{tab:results2} shows the training time of the networks without and with method \ourmethod; in all the cases our method improved the training speed by 150\%.

\begin{table}[ht!]
    \centering{\footnotesize\sffamily
    \begin{tabular}{crrr}
    \hline
    \textbf{Train Set} &\textbf{Base} & \textbf{\ourmethod} &\textbf{Speed-up}\\
    \hline
    ACDC  &  21.0s & 11.6s & 1.81x \\
    M\&Ms &  93.2s & 37.6s & 2.48x \\
    LVSC  & 254.7s & 95.1s & 2.68x \\
    all   & 427.2s & 170.3s & 2.51x \\
    \hline
    \end{tabular}}
    \caption{Mean training time of 30 epochs for CNN segmentation without \ourmethod (Base) and after \ourmethod. In all the cases, method \ourmethod accelerates the training speed; this is because the CNN processes only the data in the region of interest.}
    \label{tab:results2}
\end{table}

\section{Discussion}

Based on our results, the main achievement of method \ourmethod is the ability to identify and crop the ROIs with a very small error (Recall = 99.69\%). Furthermore, by cropping the ROIs before the CNN training, we observed that the training process was remarkably accelerated, increasing the training speed by 150\%. When considering metric DSC, we observed significant improvements when considering the entire bundle of experiments, as presented in Table \ref{tab:results} but, for dataset ACDC when used for training, the results were not promising.

\begin{figure}[ht!]
    \centering
    \includegraphics[width=.9\columnwidth]{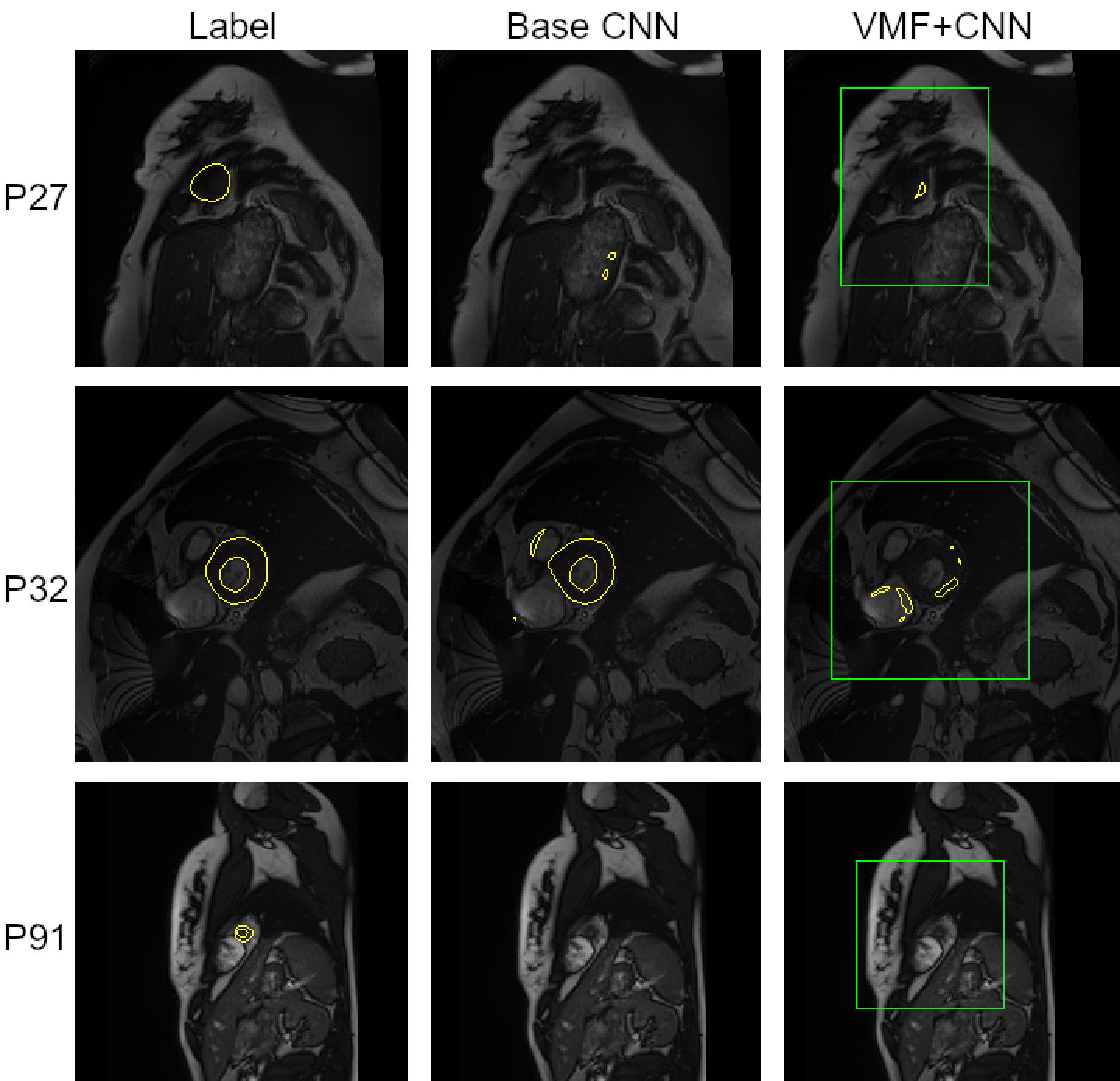}
    \caption{Examples of slices with DSC$<$0.2 found in ACDC patients number 27, 32 and 91 (rows). Left column shows the original label for each patient. The center column shows the prediction by the Base CNN. The right column shows both the output of \ourmethod (bright boxes) which is used for cropping, and the prediction of the CNN after \ourmethod.}
    \label{fig:acdc-cases}
\end{figure}

This inefficiency with ACDC is possibly related to imaging artifacts and standard operating procedures used to label the ACDC images (refer to Figure \ref{fig:acdc-cases}). In general, \ourmethod is able to automatically focus on the correct region and guide the segmentation CNN (Figure \ref{fig:acdc-cases}, P27). However, in some cases the CNN lost track of the LV when fed with a slightly displaced ROI (P32). In the rarest of cases the CNN is unable to detect anything, even when fed with a centered ROI (P91), and this is a known problem when segmenting the cardiac apex. These findings should be investigated in future work.

Most of these issues were subsequently reviewed in M\&Ms \cite{saber2020multi} by: a) complete coverage, including papillary muscles, of RV and LV; b) no interpolation of masks at the base; c) guarantees of larger RV surface in end-diastole compared to end-systole, excluding the pulmonary artery; d) sampling data from several hospitals, different patients and different equipment vendors.
The outcome of these corrections is clear when comparing the results from training in ACDC and testing M\&Ms (-20.2) vs training on M\&Ms and testing on ACDC (+6.2). M\&Ms is very likely to contain images following the same patterns of ACDC and others, but not the opposite.

Overall, method \ourmethod demonstrated significant improvements in the task of ROI detection; from the results, it became evident that the performance depends on the training dataset and on the network model -- image quality, format and intensities should be similar across datasets, and labelling standard operating procedures should be compatible. 

\section{Conclusions}
\label{sec:conclusion}
In this paper, we proposed method \ourmethod, a novel approach based on convolution operations and on the use of a Radial Basis Function to detect the ROIs in cardiac magnetic resonance images. We validated \ourmethod with a U-net CNN comparing our results to those of the canonical U-net and of the \ourmethod-CNN in three public reference datasets. According to our results, \ourmethod was able to crop 99.69\% (Recall metric) of the ROI voxels in all the datasets, being suitable to preprocess the data for CNN segmentation. \ourmethod accelerated the training process by 150\%, and also increased metric Sørensen-Dice Coefficient in the majority of our test cases ($p<.001$). Further improvement is possible by extended preprocessing of the training datasets, and change the U-net CNN after the ROI proposal step to a more advanced architecture with fine-tuning to the dataset specifics, specially in cases such as ACDC. 

As future work, we intend to expand this pipeline considering possibilities: (1) embed this method as the first layer of a CNN, so that the parameters are adjusted automatically; (2) use other networks in the segmentation step, such as 3D Unet or Feature Pyramid Networks;
(3) experiment with more datasets, preprocessing, and augmentation methods; (4) test different feature extractors for the static visual features, such as Gray-Level Co-occurrence Matrices and Gabor filters, other motion estimation methods, and Radial Basis Functions.

\section*{Acknowledgment}
We thank Dr. Diego Cardenas and Dr. Thiago Costa for reviewing equations, Nvidia Corporation for donating the GPU used in this work, and USP, FAPESP, CAPES, CNPq, Zerbini Foundation and FoxConn for supporting this research.

\section*{Declarations}
\noindent\textbf{Conflict of interest}: The authors declare that they have no conflict of interest.

\noindent\textbf{Research data}: The datasets used in this research are available upon registration on \href{https://www.ub.edu/mnms/}{https://www.ub.edu/mnms/}, \href{https://acdc.creatis.insa-lyon.fr/description/databases.html}{https://acdc.creatis.insa-lyon.fr/description/databases.html} and \href{https://www.cardiacatlas.org/challenges/lv-segmentation-challenge/}{https://www.cardiacatlas.org/challenges/lv-segmentation-challenge/}

\bibliographystyle{abbrvnat}
\bibliography{references}
\end{document}